\documentclass{article}
\usepackage[T1]{fontenc} % add special characters (e.g., umlaute)
\usepackage[utf8]{inputenc} % set utf-8 as default input encoding
\usepackage{ismir,amsmath,cite,url,amsfonts}
\usepackage{graphicx}
\usepackage{color}
\usepackage{subcaption}
\usepackage{booktabs}
\usepackage{lineno}
\title{Contrastive learning for cross-modal artist retrieval}

\multauthor
{Andres Ferraro \hspace{1cm} Jaehun Kim \hspace{1cm} Sergio Oramas} { \bfseries{Andreas Ehmann \hspace{1cm} Fabien Gouyon \hspace{1cm}}\\
   Pandora-SiriusXM, Oakland\\
   {\tt\small andres.ferraro@siriusxm.com}
}
\def\authorname{A. Ferraro, J. Kim, S. Oramas, A. Ehmann and F. Gouyon}

\begin{document}

\maketitle
\begin{abstract}
Music retrieval and recommendation applications often rely on content features encoded as embeddings, which provide vector representations of items in a music 
dataset.
%catalog. 
Numerous complementary embeddings can be derived from processing items originally represented in 
%a given modality, 
several modalities,
e.g., audio signals, user interaction data, or editorial data. 
However, data of any given modality might not be available for all items in 
%the music catalog. 
any music dataset.
%For instance, it is quite common to have access to collaborative filtering data for a limited subset of items --only the most popular ones, while audio data may be available for all items.
In this work, we propose a method based on contrastive learning to combine embeddings from multiple modalities 
%exploring 
and explore
the impact 
%--when evaluated in terms of ranking performance and inequality of retrieved items-- 
of the presence or absence of embeddings from diverse modalities in an artist similarity task.  
Experiments on 
%a public and a private dataset
two datasets 
suggest that our contrastive method outperforms single-modality embeddings and baseline algorithms for combining modalities, both in terms of artist retrieval accuracy and coverage.
Improvements with respect to other methods are particularly significant  
for less popular query artists.
%In addition, we
We
demonstrate our method successfully combines complementary information from diverse modalities, 
%while leveraging the best modality information available.
and is more robust to missing modality data (i.e., it better handles the retrieval of artists with different modality embeddings than the query artist's).
%We also show our method to yield improvements that are particularly significant for less popular query artists.
%\fabien{almost done with rewording the abstract...}
%allowing to cover 
%covering more artists by flattening the distribution of retrieved artists. 

\end{abstract}
\section{Introduction and related work}\label{sec:introduction}

% 1 page
% Outline:
% - general context
% - what problem are we trying to solve?
% - why is this an important problem for Ismir?
%        - why we focus on artists similarity?
%        - why we want to cover all the catalog?
% - why current solutions are not sufficient?
%        - maybe motivate with some fairness issues?
% - what do we propose?
%        - how do we address these problems?
% - which are the contributions of this work

The MIR community has dedicated significant effort to defining and computing music similarity in the last $20$~years. Music similarity can be used in multiple downstream tasks, from playlist continuation, music visualization/navigation, music categorization for organizing catalogs, or for 
%other 
personalized recommendations. 
The notion of similarity 
%may be 
is
subjective and
%, therefore, 
there is no consensus on how to define and evaluate it~\cite{ellis2002quest}.  To evaluate 
the performance of a music similarity algorithm, some previous works either focus on content-based aspects, % to define similarities, 
such as melody or harmony. Other works measure similarity based on \textit{cultural aspects}, such as based on the co-occurrence of items in playlists 
%or webpages,  
or on editorial data
%-- which 
--this is the 
%focus 
approach
of 
our work.%., we are taking the cultural aspects to evaluate the performance of multiple methods. %following the approach applied by~\cite{korzeniowski2022artist}.

%In the MIR community, 
Multiple methods 
%were 
have been 
proposed to compute music similarity based on a variety of data types 
%(i.e., modalities) 
related to the music, e.g., based on audio descriptors~\cite{pohle2009rhythm}, document similarity~\cite{schedl2014harvesting}, or graphs of musical connections~\cite{oramas2015semantic, korzeniowski2022artist}. 
%Additionally, in recent years multiple works propose ways to learn representations mapping from different input data to an embedding space that can be used for multiple downstream tasks~\cite{radford2015unsupervised}. For example, using large pre-trained models to obtain embeddings of audio tracks~\cite{alonso2020tensorflow} or artists' biographies~\cite{oramas2015semantic} is now a common practice and showed high generalization capabilities. 
%
%At the same time, some novel works 
Some relatively recent works
propose ways to produce embeddings --that can be used to compute music similarity-- in a supervised or unsupervised way, by training models on large amounts of data (such as audio, text or image). Such pre-trained models, which are often released publicly, may produce feature representations --i.e. embeddings-- that are effective for previously unseen tasks. %Those representations 
Such embeddings
can be computed from diverse types of modalities related to music such as audio~\cite{mccallum2022supervised, alonso2020tensorflow, alonso2022music}, tags~\cite{dieleman2011audio}, album covers images~\cite{oramas2018multimodal}, or %artist
biographies~\cite{oramas2015semantic}.
%
% need for multiple modalities to describe a song:
%
The multiple modalities of data that can describe a music item --such as audio, tags, or listening interactions-- may contain \emph{complementary} information. 
%For example, a system in some cases may be more accurate at describing the songs' tags using audio, and in other cases it may be more accurate using collaborative information~\cite{won2021multimodal}. 
For example, 
the quality and scale of audio vs collaborative data has been shown to have significant influence in autotagging tasks
%we can see a higher performance on describing the songs' tags using collaborative information or with audio, depending on multiple factors such as 
\cite{won2021multimodal}. 
% Combining different sources and types of data was particularly identified as promising for mitigating limitations and issues in music recommendation research~\cite{ferraro2019bias}. This calls for the need of combining such modalities in order to obtain a more informed representation of the music item that can perform better when considering all the possible downstream tasks and contexts.  
It therefore appears beneficial to %try to 
combine diverse complementary modalities %in order 
to obtain a more informative representation of music items.
In fact, recent research identifies the combination of diverse sources of data as specially %particularly 
promising for mitigating limitations and issues in music recommendation research~\cite{ferraro2019bias}.

Another aspect to take into account is that 
%each of the modalities related to the music may be available for only a limited set of artists or tracks. 
in any given music dataset,
data of diverse modalities might be available for %disconnected 
different subsets of items.
%Therefore, when using a given modality to retrieve similar artists, the maximum coverage of the catalog that we can use is limited only to the ones that have such information. 
Therefore, when querying with an item represented in a given modality, the maximum coverage for retrieval is limited to items for which that same modality is available, leaving out a potentially significant --and relevant-- part of the dataset.
For example, the availability of listening interactions or users' explicit feedback is highly dependent on item popularity. Therefore, for artists with very little listening and user feedback, it may not be possible to obtain  embeddings from that modality. %Additionally, modalities may be limited to the populations that can recommend/cover, either because there is no data available to produce an embedding or because the quality of information available for some artists or items is very low. 
Embeddings from other modalities may suffer from the same issue, either because there is no data available to produce an embedding or because the quality of the available information is very low. 
%An example of low-quality information may be regarding artists' tag annotations, since we may only have a single tag for some artists, the output that a tag-based model will produce is not informative. Such a problem is particularly common for less popular and new emerging artists, for which the information available is more limited. 
%We can see this issue 
For instance
in the case of a model trained on tag annotations to produce artist embeddings, where the output embedding may not be very informative for those artists that have a single or few tag annotations.  Such issues are particularly common and problematic emerging or more underground artists, for which the available information  is more limited. 

In order to mitigate 
%reduce 
the issue of availability of some modalities, it is important to combine and take full advantage of all information available so that when querying with an artist that has only one modality available, we can also retrieve artists for which we have a different modality information.
%\fabien{We should say something here about the issue of silos (see comment in the tex file below).}
%another (less obvious) issue in this scenario without a multimodal model: When querying with one specific model (say audio-only embedding -model (1)), retrieved items may be biased towards items that have only that modality, in opposition to items that have that modality and some other, i.e. in my example, biased towards cluster (1), audio-only, and artificially ranking down items from clusters (4), (7) or (5) that have audio and another modality (of course it can't reach items from clusters (2), (6) and (3), but that's less interesting because obvious).
Therefore, the 
%The 
focus of this work is to combine 
%such 
diverse modalities into a common shared space that is beneficial for 1) leveraging each modality information from the artists, and 2) allowing to operate on a single space that covers the full population of artists, ensuring that %all artists that contain some information are potential retrieval candidates for any query artist.
%any artist is a potential nearest neighbour of any other artist, regardless of the number of modalities available. 
whether or not an artist is retrieved for another does not depend on the number of modalities available. 

The problem of combining embeddings from diverse modalities in a shared representation has received some attention in the last few years. %, specially in the computer vision domain~\cite{}. 
In the music domain, there have been some works on combining embeddings by simple concatenation~\cite{oramas2017multi} or predicting one modality from another~\cite{van2013deep}. Contrastive learning techniques go beyond simple concatenation or prediction, trying to learn a shared representation between embeddings from different modalities. Some examples of research related to multimodal contrastive learning can be found in~\cite{oramas2018multimodal}, where embeddings from a shared multimodal space are used as additional features for classification, or in~\cite{huang2022mulan,manco2022contrastive} where, e.g., music audio can be retrieved from natural language descriptions. In this work, we propose to apply a contrastive learning method that maps embeddings from diverse modalities 
to a shared embedding space, 
%from different modality embeddings, 
%to improve artists' coverage, that is 
%allowing to extend 
extending the advantages of multiple modalities to populations that would not be covered otherwise.

%Combining multimodal embeddings in a common/shared representation of the artists received less attention from the community~\cite{mcfee2009heterogeneous}. In particular, there is less research dedicated to leveraging the different modalities to extend the capabilities for a larger population of artists in the catalog. In this work, we propose to apply a contrastive learning method --that already proved to be useful in multiple works~\cite{oramas2018multimodal, lee2020metric, manco2022contrastive, alonso2022music}-- to improve artists' coverage, that is to extend the advantages of multiple modalities to populations that would not be recommended/covered otherwise.

 % what is new?
In summary, in this work we propose an approach to combine the multiple encoders of a contrastive learning method, showcasing several improvements over baselines and single-modality approaches in an artist similarity task. We show under two different contexts --using an open and an in-house dataset-- that our proposed approach:
\vspace{-6pt}
%\begin{enumerate}
\begin{itemize}
    \item achieves higher performance in terms of accuracy and coverage of retrieved artists
    (\S~\ref{sec:res-perf}),
%    \item allows to cover more artists by flattening the distribution of retrieved artists (\S~\ref{sec:res-perf}), 
\vspace{-6pt}
%    \item leverages the best modality information available    
\item successfully combines complementary information from diverse modalities
(\S~\ref{sec:res-lev}),
\vspace{-6pt}
    \item is more robust to missing modality data
%    from the diverse modalities, 
    (\S~\ref{sec:robust}),
\vspace{-6pt}
    \item particularly increases the performance for less popular query artists
    (\S~\ref{sec:res-pop}).
%\end{enumerate}
\end{itemize}

\section{Methodology}\label{sec:typeset_text}
% 1 page
% - Describe our approach
% - Describe what we are evaluating
% - Describe the data sources and the different modalities
% - what data (modalities) are we using
% - How is done dataset split
% - metrics used
% - method used
% - Preliminary analysis of the output embeddings

%In this section, we describe the proposed method and baselines applied, the experiments and metrics that we used to evaluate them, and the data used to train the models.

\subsection{Single-Modality Embeddings and Contrastive Method}\label{sec:contrastive}

% - pre-trained models
% - some models are publicly available some we had to train
% - we train them in a larger set of data
% - we use cosine similarity between the embeddings to rank artists
% - the data used to train these models

In this work, 
%we combine multiple sources that encode information from diverse modalities that are related to the artists. 
we use three modalities, namely: tags, user-listening interactions (i.e. collaborative filtering data, referred to as CF), and audio information. In all cases, we use pre-trained models to obtain embeddings for each of the modalities. We evaluate artist similarity performance using the embeddings from the pre-trained models directly, and compare to the performance when using the embeddings produced by our contrastive method which is trained with the same embeddings from pre-trained models. 

In these experiments we apply a contrastive learning loss based on InfoNCE~\cite{oord2018representation}. 
Specifically, we define the contrastive loss between two modalities, $\pmb{\psi}_{a}$ and $\pmb{\psi}_{b}$, as:
\begin{math}
% \mathcal{L}_{\pmb{\psi}_{\alpha},\pmb{\psi}_{b}} = \sum\limits_{i=1}^{M}-\log\frac{\Xi(\pmb{\psi}^{i}_{\alpha}, \pmb{\psi}^{i}_{b}, \tau)}{\sum\limits_{k=1}^{2M}\mathbbm{1}_{[k\neq i]}\Xi(\pmb{\psi}^{i}_{\alpha}, \pmb{\zeta}^{k}, \tau)}
\mathcal{L}_{\pmb{\psi}_{a},\pmb{\psi}_{b}} = \sum\limits_{i=1}^{M}-\log\frac{\Xi(\pmb{\psi}^{i}_{a}, \pmb{\psi}^{i}_{b}, \tau)}{\sum\limits_{k=1}^{2M}\mathbb{1}_{[k\neq i]}\Xi(\pmb{\psi}^{i}_{a}, \pmb{\zeta}^{k}, \tau)}
\end{math}  
\noindent
, where M is the batch size and $\tau$ is the temperature parameter. We define $\Xi(\mathbf{a}, \mathbf{b}, \tau) = \exp(\text{cos}(\mathbf{a}, \mathbf{b})\tau^{-1})$, based on the cosine similarity. $\pmb{\zeta}^{k}$ is defined as $\pmb{\psi}^{k}_{a}\text{, if }k\leq M$ and else $\pmb{\psi}^{k-M}_{b}$. 
This loss function attempts to minimize the distance between the 
%representations
modalities
of the same artist while maximizing the distance 
%between any representations
with any
modality
from other artists.

We use three encoders --one for each modality-- that will produce three representations in our shared space for each artist. During training\footnote{For both datasets we use Adam optimization with a learning rate of 0.0001 and temperature of 0.1. We use a fully connected layer of 256 for the CF encoder, two layers with 512 and 256 for the Audio encoder and 4 attention heads of 256 for the tag encoder. The learned space has 200 dimentions. Batch size for $C_{OWN}$ is 2048 and for $C_{MSD}$ is 128.} we minimize the sum of the pairwise losses between each of the modalities as in \cite{ferraro2021enriched}: $\mathcal{L}_{\text{tot}} = \mathcal{L}_{\text{Audio-Tag}} + \mathcal{L}_{\text{Audio-CF}} + \mathcal{L}_{\text{Tag-CF}}$

Once the model is trained with the contrastive method and we want to use it for inference, for a given artist, we aggregate the output of each internal encoder by averaging all available information.% as shown in Figure~\ref{fig:contr_diag}.

%\begin{figure}
% \centerline{
% \includegraphics[trim=0 35 0 35, clip, width=0.80\columnwidth]{figs/contr_diag.pdf}}
% \caption{Illustration of the inference process with the contrastive method. The output of three encoders (audio-based, tag-based, and cf-based) are averaged to produce the final representation for artist a1}
%\label{fig:contr_diag}
%\end{figure}

\subsection{Training Data}

In order to investigate the effectiveness of our contrastive method under different situations, we 
%decided to 
train our model using two independent datasets: We use a dataset based on public data to facilitate the reproducibility of some of the results.
And we also 
%built and apply 
use
an in-house dataset that contains multimodal information for a larger set of artists.

Training our model 
%with the contrastive method, it is required to have the 
requires \emph{full coverage of the three modalities for all artists} --tag-based embeddings, CF embeddings, and audio embeddings. 
%\fabien{The sentence above is correct, right Andres?}
%Therefore, to build the public dataset used to train the models, 
For the public dataset, we use the Million Song Dataset (MSD)~\cite{Bertin-Mahieux2011} and its connections with other datasets to collect tags, audio and CF embeddings. We collected audio track embeddings using the public unsupervised model from~\cite{mccallum2022supervised} to extract embeddings from MSD audio previews, then we averaged all audio tracks embeddings for each artist. The tagging data was collected from the MSD500 dataset~\cite{won2021multimodal} and embeddings were computed using PMI factorization~\cite{oramas2017multi} of 500 tags. The CF embeddings were obtained using weighted matrix factorization~\cite{DBLP:conf/icdm/HuKV08} based on the Echonest Profile dataset,\footnote{Specifically, we aggregated the per-song listening counts corresponding artists such that we obtain the `user-artist' listening matrix.} with Gaussian process-based Bayesian hyperparameter tuning~\cite{tim_head_2018_1207017}. We gathered information from the three modalities for  $17,478$ artists.%\footnote{\url{http://github.com/PandoraMedia/ismir23-artist} }

For the in-house dataset (hereafter, OWN) we collected tags, CF, and audio information for $38,301$ artists. This 
%larger 
dataset 
is larger than MSD and includes what we believe is \emph{higher-quality tags and CF data},
which
allows us to compare the performance of our approach in a different setting. 
%on a subset of our catalog. 
The CF information is computed from very large amounts of user-listening interactions on a 
%private 
%music 
streaming
platform. The audio embeddings are computed using the supervised model\footnote{i.e. a \emph{different} model for audio embeddings than when training on MSD.} described in \cite{mccallum2022supervised}. The tag embeddings are computed using PMI factorization from a total of $6,421$ different tags, which are a combination of manual and automatic annotations. Since our pre-trained models for audio and CF are at the track level, 
we compute artist embeddings by averaging over artist track embeddings. 

In the remainder of this work, we refer to the model trained with the contrastive method with in-house data as $C_{OWN}$ and the model trained with public data as $C_{MSD}$. % Add the number of artists used to train the model

\subsection{Evaluation Dataset}\label{sec:eval-data}

%In this paper, t
The ground truth 
%used 
for artist similarity is defined herein by the OLGA public dataset~\cite{korzeniowski2021artist}, containing artist similarity information collected from AllMusic. Our evaluations are therefore based on a \textit{cultural} ground-truth, following~\cite{korzeniowski2022artist}.

We collected data from the MSD dataset for the original $17,646$ artists in OLGA. We obtained tag data from the MSD500 for $10,971$ ($62$\%) artists, user interaction data from the Echonest Profile dataset for $15,389$ ($87$\%) artists, and audio embeddings using MSD audio previews for $100$\% of the artists.\footnote{Note that
we don't control for artist separation between MSD, OWN and OLGA. But even if some artists may be present in both train and test sets, the artist similarity information from OLGA 
%--collected from Allmusic-- 
is \emph{only} used for evaluation, and is never used during the training of the 
%pre-trained 
single-modality embeddings
nor the contrastive models on either MSD or OWN. }

We also create a subset of OLGA where \emph{all} artists contain complete tags, user interaction, and audio information from MSD. We refer to this subset as OLGA Full Modality Coverage (FMC), which contains $9,474$ artists and it is also mapped to our internal dataset. 
The OLGA FMC subset is used to 
%meaningfully 
compare the results of multiple methods pre-trained on different and independent datasets.

\subsection{Evaluation Conditions}

In order to provide 
%more 
insights on the performance of the contrastive method, we conduct analyses under 3 different situations, varying the degree of availability 
%from 
of
the different modalities in the evaluation data:

\vspace{3pt}
\noindent\textbf{Raw evaluation dataset:}
In one condition, we compare the methods using all the artists in the OLGA dataset. In this case, we are interested to understand 
%how is the 
performance 
in a scenario of a real --uncontrolled-- evaluation dataset, accounting for some organic imbalance of the availability of data in different modalities.
%when we do not have an intervention on how much information is available from the different modalities. 

\vspace{3pt}
\noindent\textbf{Full Modality Coverage:} In another condition, we use the OLGA FMC subset where all artists contain CF, tags, and audio embeddings in both MSD and OWN datasets. In this case, we want to 
%analyze the performance of the method 
understand performance
%without being affected by the availability of the information.
while factoring out the potential influence of one or another modality being only partially available in evaluation.
 
\vspace{3pt}
\noindent\textbf{Systematic variation of modality coverage:}\label{modalitygroups}
We also perform multiple comparisons by grouping artists from OLGA depending on how many modalities are available. Here, we want to look at how much the contrastive method and the baselines are capable of doing cross-modality retrieval when using different modalities as input. In particular, 
we want to see whether or not they are capable of retrieving artists that have different modality information available compared to the query artists. 
%we want to test the capability to retrieve artists which available modalities differ from the query artists'. 
Therefore, in this part, we create $7$ groups of artists --at random-- of equal size with each group containing one, two, or three modalities (namely, CF, audio, tag, CF+audio, CF+tag, audio+tag, audio+CF+tag).
We refer to these groups as `Modality Groups'. 
It is important to highlight the artificiality of this setting. We are considering an extreme case only to evaluate cross-modality retrieval capabilities of the methods. 
%the possibility that the method gives of retrieving items across modalities 
We are not considering here the accuracy of these results since it is already evaluated in the other analyses.
%\end{itemize}

%Our goal is to understand the effectiveness of the proposed method in identifying similar artists considering the typical restrictions of a real-world setting. We see that typically CF methods are only  available for a limited set of artists that are more popular. Then, manual tags are available for a larger set of artists. Finally, audio is usually available for the full set of artists. Following this rationale, we artificially reduce the information available from the artists in the ground truth. Consequently, we simulate as we only have CF, tags, and audio information for 1/3 of the artists (Group 1). Then we only use the tags and audio information for another 1/3 of the artists (Group 2) and only the audio for the rest 1/3 of the artists (Group 3). We evaluate the performance of each of these groups independently and also we evaluate the performance of all the groups together (Full). 

\subsection{Baseline multimodal approaches}

For multi-modal baselines, we employ two conventional models: PCA, and Gaussian random projection~\cite{DBLP:conf/kdd/BinghamM01, Johnson1984}~(which we refer to as Rand).\footnote{For both algorithms, we employ the  standard implementation provided from \texttt{scikit-learn}~\cite{scikit-learn}.} %\jaehun{
For fitting these models, we consider artists who have access to all modalities. Their multimodal embeddings are concatenated and treated as a single feature vector.
%}
It yields a dimensionality of $2,063$ for the MSD dataset, and $2,528$ for the OWN dataset. We set the reduced dimensionality to $200$, which is the same size as the %final 
embeddings of the contrastive model. %\jaehun{
If an artist has a missing modality in the prediction phase, we employ the global mean embedding of the missing modality.\footnote{This does not happen in FMC}% to be concatenated with the embeddings of existing modalities.
%}\fabien{How are the embeddings computed in cases of missing data?}\jaehun{please check the added description above.}

\subsection{Metrics}
%Maybe add a subsection named {NDCG and Gini}?
\vspace{3pt}
\noindent\textbf{Accuracy:}
%We use two commonly-used metrics to understand the effectiveness of the methods.%We use several metrics usually used to evaluate accuracy in information retreival and recommendation lists. We use precision and recall at multiple cutoffs to see how accurate the retrieved artists are according to the ground truth. We also consider r\_precision, which adapts the cutoff for each artists to the number of similar artists in the ground truth.
We consider nDCG@200 to measure how accurate the retrieved artists are compared to the ground truth while taking into account the position in the ranking of the retrieved artists, a metric considered 
%less sensitive 
robust
to missing relevance information~\cite{valcarce2020assessing}.\footnote{We focus on nDCG@200 in this work, as we experimentally observed high correlation with other retrieval metrics such as precision, recall, and R-Precision.}
%less sensitive to missing label~\cite{valcarce2020assessing}.%\jaehun{
%}

\vspace{3pt}
\noindent\textbf{Distribution:}
We also compute the Gini@200 index, measuring the distribution of the top 200 retrieved artists in each experimental condition across the whole set of artists. A lower value of Gini indicates that the recommendations across artists are more uniformly distributed --covering more artists retrieved-- while a higher value of Gini indicates that the recommendations are focused  on only the few same artists.
%We also consider the measure of coverage, which indicates what is the proportion of the catalog that is recommended at least once. \fabien{This is not accurate anymore, is it?}

\vspace{3pt}
\noindent 
We compute the confidence interval using the bootstrap method~\cite{DBLP:books/sp/EfronT93} on the evaluation artist population. We report them in Figure \ref{fig:total_performance} at 95\% confidence level.

%\subsubsection{Expected Contrastive Loss}
\vspace{3pt}
\noindent\textbf{Expected Contrastive Loss:}
We propose an additional metric that we named Expected Contrastive Loss (ECL). We use this measure to analyze to what extent an artist is coherent with respect to their multimodal representations. 
From how we defined the loss in Section~\ref{sec:contrastive}, %the contrastive method attempts to minimize the distance among the set of internal embeddings obtained from diverse modalities belonging to a target artist, while maximizing the distance between any embeddings from other artists. Thus, 
a high loss value implies that the artist is relatively difficult to be distinguished from other artists.
%, and vice versa for the artist with a lower loss value. 
Once the training is reasonably progressed, we employ ECL to quantify how ``coherent'' the artist is with respect to their internal representations obtained from the different modalities, which is defined as:
% TODO: Math notation is too much abused here. should be fixed before the final submission, but maybe lower priority if other things should be done
\begin{math}
    ECL(i, u, v) = d_{ii}^{uv} - \mathbb{E}_{j\setminus i}[d_{ij}^{uv}]\text{,}
\end{math}
where $i$ and $j$ denote artist index, while $u$ and $v$ refer to the modality index. $d_{ij}^{uv}$ means the cosine distance between artist $i$ from modality $u$ and artists $j$ from modality $v$. Taking expectation over all the possible modality pairs leads to the final coherency measure for artist $i$: $ECL(i) = \mathbb{E}_{u, v\setminus u}[ECL(i, u, v)]$.

\vspace{3pt}
\noindent\textbf{Clustering:}
We further analyze the multimodal embedding space of the %trained 
contrastive model, by investigating how well the artist embeddings are clustered. The contrastive method essentially can be seen as a  ``supervised'' clustering task, where we %want to 
minimize the distance among ``positive points'' (i.e., multimodal embeddings from an artist) and maximize the distance between those to the ``negative points'' (i.e., embeddings belonging to the other artists). It implies that an artist will get a higher training loss when the embeddings %of an artist 
are dispersed and overlapped with the embedding cluster of other artists, while the opposite cases will get lower values. The model will fit the multimodal embedding space such that the artist embeddings poorly clustered initially have more concentrated and distant clusters. While the contrastive learning implements this naturally by its loss function, there are other well-known measures for the validation of the clustering methods, such as \textit{intra-cluster distance ($CD_{intra}$)} 
indicating how an artist embeddings are well clustered together, 
and \textit{inter-cluster distance ($CD_{inter}$)} indicating how an artist-specific embedding cluster is 
far and 
distinct from others'.\footnote{we compute $CD_{intra}$ as the mean cosine distance between multimodal embeddings of an artist to their centroid in the multimodal space of learned contrastive model. $CD_{inter}$ is computed as the mean distance between the centroids of target artist and of all the other artists.}

\section{Results}
% 1 - 1 and 1/2 page

\subsection{Performance comparison of contrastive method}\label{sec:res-perf}
%\subsection{Artist Retrieval Accuracy and Coverage}\label{sec:res-perf}
%\subsection{Artist Retrieval Performance}\label{sec:res-perf}

%In this section, 
We now 
%compare 
look at
the performance of the contrastive method when some modality information is missing in the evaluation dataset (using the 
%full 
raw
OLGA dataset) and when all modalities are available for each artist (FMC subset). We also compare the performance of the contrastive method 
%using ndcg and gini index,
to the baseline methods and 
%using the pre-trained embeddings independently. 
to single-modality embeddings.

\subsubsection{Performance 
%of contrastive method 
with incomplete modality information}

Focusing on the different combinations of input modalities to the contrastive method, 
we can see 
in Table~\ref{tab:full-models}
that the highest nDCG result is obtained when combining all modalities as input.  We therefore focus only on this model for the remainder of the work. %More detailed results are omitted for space limitations.

\begin{figure}%
\centering
\begin{subfigure}{.355\columnwidth}
 \includegraphics[trim={0 0.5cm 9cm 0.5cm},clip,width=\columnwidth]{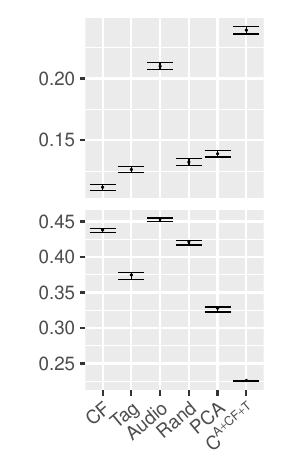}%
 \caption{MSD-OLGA}
 \label{fig:perf-olga}
\end{subfigure}%
\begin{subfigure}{.354\columnwidth}
\includegraphics[trim={4.5cm 0.5cm 4.5cm 0.5cm},clip,width=\columnwidth]{figs/performance_ndcg_gini_alt.pdf}%
 \caption{MSD-FMC}
 \label{fig:perf-fmc-msd}
\end{subfigure}%
\begin{subfigure}{.31\columnwidth}
\includegraphics[trim={9.6cm 0.5cm 0cm 0.5cm},clip,width=\columnwidth]{figs/performance_ndcg_gini_alt.pdf}%
 \caption{OWN-FMC}
 \label{fig:perf-fmc-own}
\end{subfigure}%

 \caption{Performance comparison between contrastive and other methods. Training with MSD (a and b) or with OWN (c), Evaluation on OLGA (a) or on FMC (b and c).}
 \label{fig:total_performance}
\end{figure}

Figure~\ref{fig:perf-olga} shows the results for all artists in OLGA. % data 
% \fabien{we need a Figure 2-a or Figure2-b here...}). 
%when some modalities have missing information. 
We can see that when using features from MSD, the contrastive method outperforms the baselines and the original embeddings in all the metrics. The contrastive method always gives a better Gini compared to %all 
the other methods --which means that the distribution of retrieved artists is more uniform-- while outperforming the other models in nDCG.\footnote{OWN-OGLA is omitted since we observe a similar behaviour.} %Thus, giving more accurate results and covering more artists. 

%If we also compare the contrastive method when using all $7$ different combinations of modalities as input (Table~\ref{tab:full-models}), we can see that the highest nDCG result is obtained when combining all modalities as input.  Therefore, we focus only on this model for the remainder of the work. %More detailed results are omitted for space limitations.

%When pre-training the models with our private data, precision is slightly higher when using the pre-trained embeddings on collaborative data, however, the contrastive method gets a comparative performance according to precision, and a better recall and ndcg.  Additionally, when looking at %coverage and 
%gini index, we can see that the contrastive method gives higher performance than all the other methods. In our private dataset, the results using contrastive method outperform the baselines when considering both ndcg and gini. % and coverage.

\begin{table}
 \begin{center}
 \footnotesize
 \begin{tabular}{|p{0.08\textwidth}|l|c|c|c|c|}
  \hline
     & \multicolumn{2}{|c|}{\bfseries OLGA}  & \multicolumn{2}{|c|}{\bfseries FMC}  \\
   & nDCG@200 & Gini & nDCG@200 & Gini  \\
  \hline
$C^{A+CF+T}$	&\textbf{0.2387}&	0.2264&\textbf{0.3560}&0.1666 \\	
$C^{A+T}$	&	0.2282&	0.2035&0.3407&0.1559\\	
$C^{A+CF}$	&	0.1381&	0.3425&0.2319&0.1873\\	
$C^{CF+T}$	&0.1781&	0.3467&0.3082&0.1917\\	
$C^{A}$	&	0.2338&	\textbf{0.1857}&0.3471&\textbf{0.1353}\\	
$C^{T}$	&	0.1232&	0.4939 &0.2554&0.1745\\	
$C^{CF}$	&	0.1381&	0.3425 &0.2319&0.1873\\	
							\hline
 \end{tabular}
\end{center}
 \caption{Evaluation of the contrastive method trained with MSD data using all combinations of modalities for OLGA dataset and FMC subset. }
 \label{tab:full-models}
\end{table}

\subsubsection{Performance 
%of contrastive method 
with complete modality information}

When we look at the results 
%of the artists that contain information for the three modalities in the OLGA dataset 
with Full Modality Coverage
(Figures~\ref{fig:perf-fmc-msd} and \ref{fig:perf-fmc-own}), %we see that the 
the contrastive method outperforms the baselines and the pre-trained models in all the metrics both when trained with MSD data or with OWN data. %This suggests that when information from all modalities is available, the contrastive method is able to accurately retrieve artists and with a better distribution. %It is important to highlight that when selecting the top 5 candidate artists, the contrastive method already covers 90\% of the catalog, which is much higher than using the baselines or the pre-trained models.

When looking at baseline performance between OLGA and FMC (Figures~\ref{fig:perf-olga} and \ref{fig:perf-fmc-msd}), we can see that in the latter, baselines are relatively close to the best single-modality embeddings, but in the former (i.e. with incomplete modality information) their performance drops significantly lower than the best single-modality embeddings. This is something we do not observe with the contrastive method, which suggests that the baseline models are more limited in the capabilities of retrieving 
artists
that miss some of the modalities from the query artist, while our contrastive method may be more robust to missing modality information. 
%in the different modalities. 
We investigate this further in Section~\ref{sec:robust}.

%By comparing the performance of the baselines between % the %full 
%OLGA %dataset 
%and %the 
%FMC % subset for MSD dataset, 
%(Figures~\ref{fig:perf-olga} and \ref{fig:perf-fmc-msd}),
%we can see that when all the modalities are available they obtain a comparatively higher nDCG than when some modalities are missing. This is something that we do not observe when using the contrastive method. This suggests that our contrastive method may be more robust to missing information in the different modalities, %which 
%we investigate this further in Section~\ref{sec:robust}.

%When pre-training the models with our private data, the gini index is %and coverage are also 
%better for the contrastive method while at the same time obtaining a higher performance according to ndcg. %, recall, and r\_precision. The r-precision for the contrastive method is 0.1679 while for the collaborative method is 0.1635. 
%In our private dataset, the results using the contrastive method also outperform the baselines when considering ndcg and gini index.% and coverage. 

%We do not observe a significant difference in the results when evaluating using the full OLGA data or using the subsample of artists that contains all three modalities. In both cases, contrastive gives a similar --or better-- performance according to precision while giving a higher performance in Gini and coverage. This suggests that contrastive is robust to different availability of data for the artists.

%\subsection{Leveraging best quality information available}\label{sec:res-lev}
\subsection{Combining complementary modality information}\label{sec:res-lev}

If we focus only on the 
%pre-trained models
single-modality approaches,
and MSD pre-training,
%in Figure~\ref{fig:total_performance}, 
Audio gives the best single-modality performance
%on all metrics
in both %the full OLGA dataset and the FMC subset 
OLGA and FMC (Figure~\ref{fig:perf-olga} and \ref{fig:perf-fmc-msd}). 
On the other hand, when pre-trained with OWN, CF is slightly better than Audio and Tag (Figure~\ref{fig:perf-fmc-own}).
These results suggest that 
%the 
performance 
%for predicting similarity 
is highly dependent on the quality of the data used to pre-train the single-modality embeddings. 
%Nevertheless, we see that contrastive is able to successfully combine complementary information from each modality and outperforms the best pre-trained model in all cases.
Results from Figure~\ref{fig:perf-fmc-msd} and \ref{fig:perf-fmc-own} also suggest that, whichever single-modality embedding is best, 
our contrastive method is able to successfully build on top of it and still gain in performance by
combining complementary information from other embeddings.

\subsection{Robustness to missing 
modality
data}\label{sec:robust}

\begin{table}
 \begin{center}
 \footnotesize
 \begin{tabular}{|p{0.04\textwidth}|l|l|l|l|l|p{0.04\textwidth}|p{0.04\textwidth}|}
 \hline
 &Audio&CF&Tag&Rand&PCA&$C_{MSD}$&$C_{OWN}$\\
  \hline
  Entropy&0.76&0.79&0.79&0.73&0.96&\textbf{1.86}&1.59\\
\hline
 \end{tabular}
\end{center}
 \caption{Entropy of each model for Modality Groups. Higher values indicate better distributed retrieved artists.}
 \label{tab:entropies}
\end{table}

\begin{figure}%
\centering
\begin{subfigure}{.5\columnwidth}
 \includegraphics[width=\columnwidth]{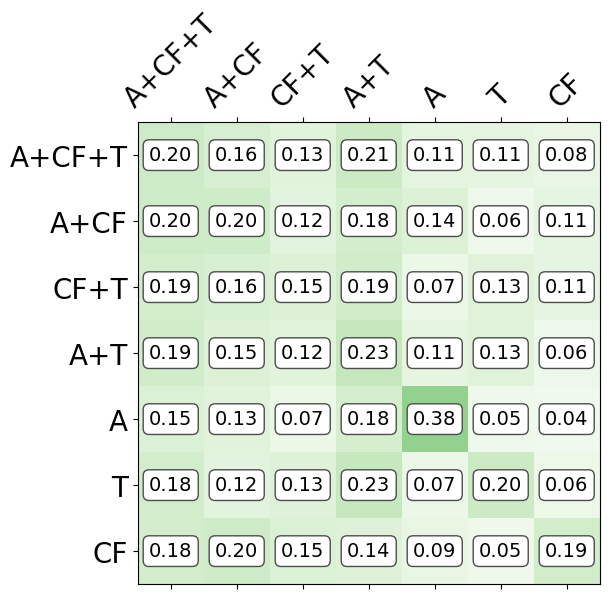}%
 \caption{Contrastive - MSD training}
 \label{fig:msd_matrix}
\end{subfigure}%
\begin{subfigure}{.5\columnwidth}
\includegraphics[width=\columnwidth]{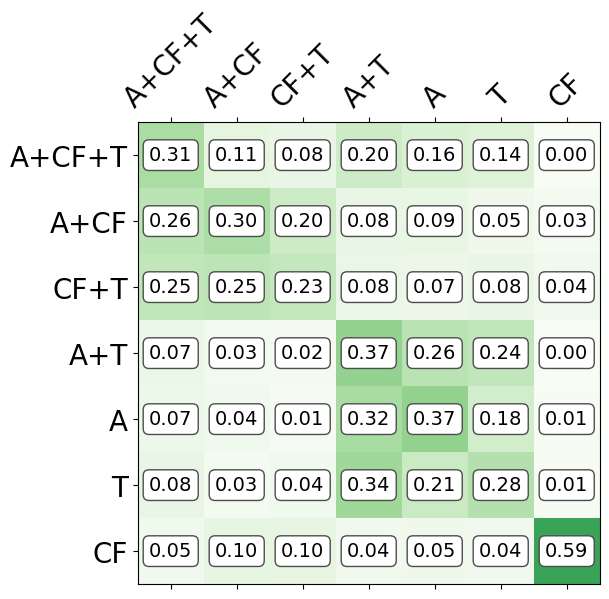}%
 \caption{Contrastive - OWN training}
 \label{fig:own_matrix}
\end{subfigure}%

\begin{subfigure}{.5\columnwidth}
 \includegraphics[width=\columnwidth]{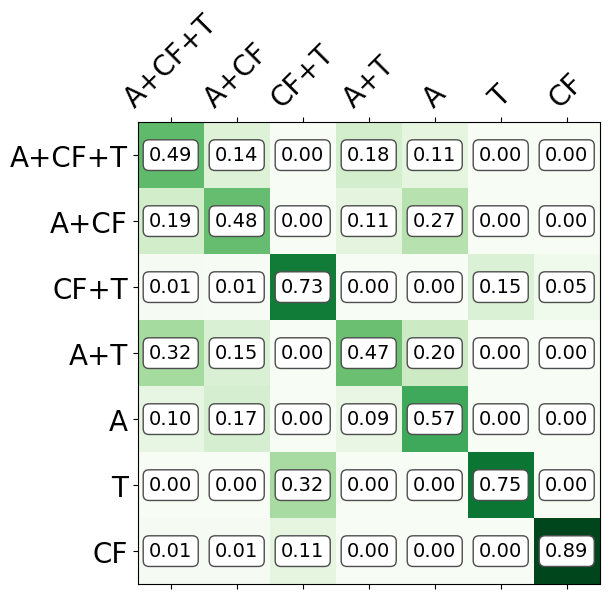}%
 \caption{PCA - MSD training}
 \label{fig:msd_pca_matrix}
\end{subfigure}%
\begin{subfigure}{.5\columnwidth}
\includegraphics[width=\columnwidth]{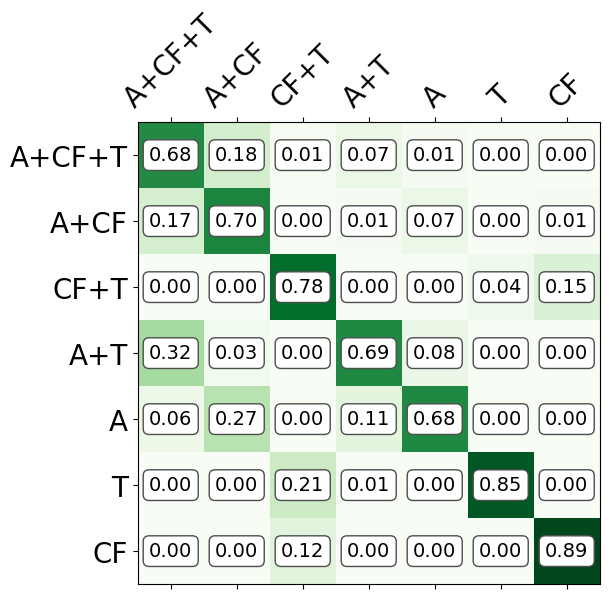}%
 \caption{PCA - OWN training}
 \label{fig:own_pca_matrix}
\end{subfigure}%
 \caption{Analysis of modality-group dependency ratio when restricting the information available for each group to one, two, or three modalities. Rows indicate the groups %of the artists 
 used to make the queries and the columns are the groups of retrieved artists. Darker green indicates a higher concentration of the retrieved artists in that cell. The color scale is normalized across all figures. Groups of artists are randomized, 
 %(see~\ref{modalitygroups}),
 so an ideal situation is a homogeneous color in the full matrix.}
 \label{fig:matrices}
\end{figure}

In this subsection, we further analyze how the contrastive method would be able to retrieve artists depending on the available information for the query artists and the 
candidates for retrieval. 
In Figure~\ref{fig:matrices}, we can see how artists are retrieved from each of the Modality Groups %~$7$ groups of query artists 
when only considering the top~$5$ results for each query artist. Typically we see that with the contrastive method, the same group used for query comprises between $15\text{-}38$\% of the retrieved artists. We see however an exception for the CF group which obtains a larger portion of the retrieved artists~($59$\%) when using OWN data to train the models. %We see that such a situation is not repeated when models are trained using MSD.

When we do a similar comparison for the PCA baseline method,  we see in Figure~\ref{fig:matrices} that there are higher percentages in the diagonal of the matrix. This indicates that most of the retrieved artists are concentrated in the same modality group 
%of artists 
used to make the query. Therefore, 
these results highlight the difficulty for the PCA baseline method to retrieve artists beyond the query artist's modality. %bias of the baseline methods to retrieve artists with only limited information available.

In Table~\ref{tab:entropies} we compare the entropy of each model for the Modality Groups. A higher entropy  indicates that retrieved artists are better distributed across the different modality groups, i.e. that retrieval is less biased by the query modality  --or more robust to partial modality data in the query. 
We can see that the
contrastive model is more robust
to missing modality data 
than 
the single-modality embeddings and the baseline approaches to combining modalities. This is true when trained with MSD or with OWN. 

\subsection{Effect of Popularity}\label{sec:res-pop}

%
% - for artist retrieval, popularity is deterministic both for learning and evaluation
% - intuitively, the more popular, we know more about those artist, which means likely we will observe better performance
% - to confirm this, we computed the total listening count per each artist as the popularity proxy measure, and compared performances and loss function.
% - In figure {fig:pop_vs_loss_vs_performance}, it is quite visible taht 
%

\begin{figure}
    \centering
    \includegraphics[trim=0 1 5 5, clip, width=0.3\textwidth]{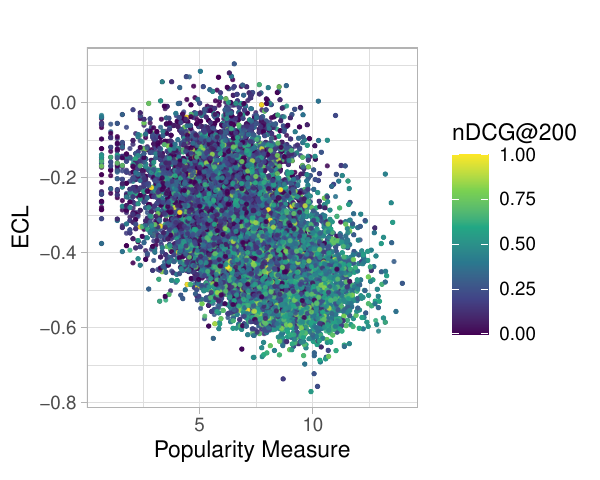}
    \caption{Scatter plot of artists based on the popularity proxy measure and the ECL. Each point represents an artist, where the color brightness represents the per-artist retrieval performance (nDCG@200). It is computed on the FMC subset with MSD data.}
    \label{fig:pop_vs_loss_vs_performance}
\end{figure}

\begin{figure}
    \centering
    \includegraphics[width=0.44\textwidth]{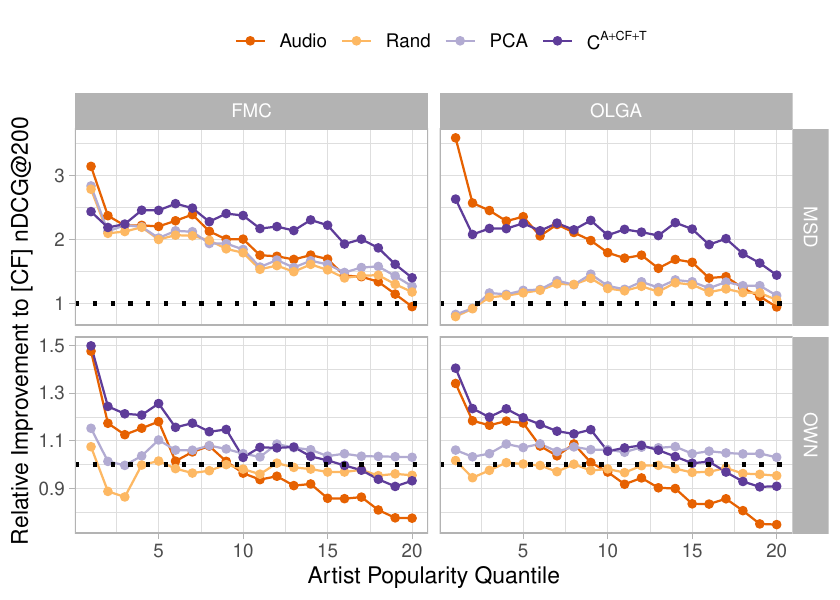}
    \caption{Relative retrieval improvement against CF modality. The $x$ axis represents the grouped popularity quantile in 20 levels, meaning the first group includes artists whose popularity is under 5\% percentile, while the top 5\% popular artists belongs to the last group. The $y$ axis is proportional improvement of nDCG@200 compared to the CF embedding model. The dotted horizontal line indicates the retrieval performance of CF modality. FMC and OLGA are evaluation datasets. MSD and OWN are training conditions.}
    \label{fig:pop_vs_performance}
\end{figure}

Artist popularity may be a deterministic factor in artist retrieval, both for 
%learning 
training
and evaluation. Intuitively, we likely have more data about popular artists, which implies more multimodal data is available for 
%learning. 
training.
At the same time, 
%for some evaluation metrics, 
the scale of evaluation metric themselves can be inflated as more popular artists would have more ground truths (annotated as `similar artists').
% To confirm this, we computed the natural log of the total listening count per each artist as a popularity proxy measure, and then further compare them to other learning and evaluation measures.
To confirm this, we compute a proxy measure for the artist popularity (POP) as $\text{POP}(\text{artist}) = \text{log}(\#\text{listen} + 1)$,\footnote{$\#\text{listen}$ denotes the total listening count of the artist, computed from the MSD-Echonest Profile dataset.} and then further compare it to other 
%learning 
training
and evaluation measures.

Firstly, we compare POP with ECL and the retrieval performance. Figure \ref{fig:pop_vs_loss_vs_performance} shows that there is correlation among POP, ECL, and nDCG. In particular, ECL has a negative correlation with nDCG. This is a desirable outcome as a model that minimizes the contrastive loss 
% therefore 
recommends ``similar'' artists even though such a model is not being explicitly shown artist-relatedness ground truth during  training. Meanwhile, POP also correlates with nDCG, which demonstrates the confounding effect of popularity to the task itself.

Further, we investigate how multimodal models interact with artists with different popularities. One of the benefits of employing multiple modalities is the potential mitigation of the information void for ``cold-start'' artists from their music audio data. For MIR applications, audio is likely accessible even when some of the other modalities are not readily available. For instance, the CF modality is not available before artists' songs are consumed by the listeners. To confirm whether the audio and further multimodal embedding models would benefit less popular artists via multimodality, we divided the artists in $20$~groups by 
%the quantiles of POP. 
popularity quantiles.
For each group, we further compute the relative improvement of retrieval performance (nDCG) compared to the CF single modality model.

Figure \ref{fig:pop_vs_performance} suggests that the original audio embedding achieves better performance for the less popular artists in all training and evaluation conditions. The contrastive model shows improvements for the majority of the groups compared to the audio, while it may have smaller or no improvement over audio in the least popular group for the MSD dataset. In the OWN dataset, a similar trend is observed where the contrastive model shows a small decline for the most popular groups compared to the original CF embeddings. The two baseline models indicate relatively flat results except in the case of the MSD-FMC subset, which implies that their prediction may be more reliant on the CF modality. For the MSD-FMC subset, both baselines follow similar trends to the audio and contrastive model.

% \subsection{Coherency/Uniqueness of Learned Artist Embeddings}
\subsection{Multimodal Embedding Space Analysis}

We conduct a correlation study of multiple measures where, for each artist, we compute clustering measures and other key 
%learning 
%training
indicators such as contrastive loss ECL, retrieval performance (nDCG@200), and finally the popularity measure. In this way, we expect to obtain a better understanding of what contrastive learning achieves in terms of clustering of embeddings, and how they are connected to retrieval performance and popularity.

\begin{figure}
    \centering
    \includegraphics[trim={10 8 10 13},clip, width=0.20\textwidth]{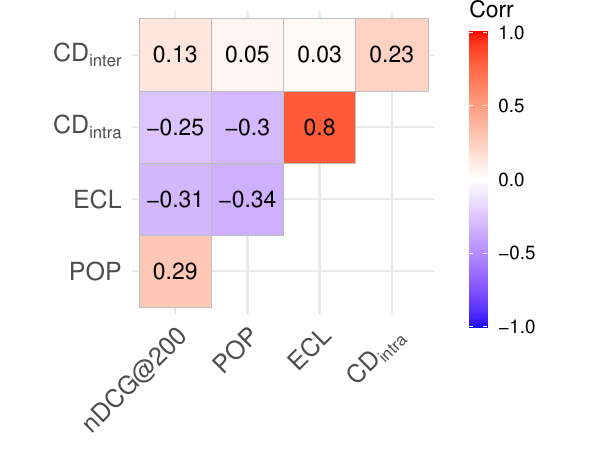}
    \caption{Correlation (Kendall's $\tau$) among variables of interest. Each cell indicates value of $\tau$ between two associated variables. POP denotes the popularity measure.}
    \label{fig:c_u_corr}
\end{figure}

The result of the correlation study can be found in Figure \ref{fig:c_u_corr}. %It shows 
We see that the ECL is highly correlated to 
%the 
$CD_{intra}$, while almost independent to $CD_{inter}$. Notably, in terms of magnitude, all other measures (ECL, $CD_{intra}$, and POP) are relatively more correlated to nDCG compared to $CD_{inter}$, and also correlated to each other.\footnote{We focus on the magnitude, as the goal of this study is to investigate the degree to which some of the key 
%learning 
%training
indicators are associated with clustering quality measures % such as $CD_{intra}$ and $CD_{inter}$, 
in absolute manner}

% It suggests that the contrastive learning more focus on tightening the per-artist embeddings than making each artist cluster spread.
These relations suggest that our contrastive learning method aims at producing an artist embedding space where the diverse modalities of an artist occupy a coherent region, but not necessarily a region that is unique to the artist.
$CD_{inter}$ shows lower correlation with most of the other measures, which confirms its relatively small connection to the contrastive learning and the artist retrieval downstream task.
% We hypothesize that the similarity structure provided by the multi-modality may act as a constraint that prevent maximizing $CD_{inter}$.
We hypothesize that this is because the maximization of $CD_{inter}$ is constrained by the artist similarity inherent in the multimodal information and ultimately preserved.
This is desirable if the ultimate goal is a representation that can measure artist similarity.

%\section{Discussion}
% 1/2 -1 page

%- Contrastive always give higher Gini index which means ...
%Looking at the precision:
%- Depending on MSD or OWN we see that CF may perform better or audio embeddings depending on the quality of these embeddings.
%- Overall it seems that CF trained in own outperforms the others. This is expected since evaluation 
%- in full evaluation data, is audio that performs better while in each independent group is the CF.
%- this is aligned with our private dataset used for evaluation, here audio get slighly better results. And this dataset is not artificial, so we it supports the idea that audio models perform well when considering a large population of artists and that contrastive always allows a better distribution (and coverage) of the recommendation.
\section{conclusion and future work}
% 1/4 - 1/2  page
In this work, we propose a method based on contrastive learning to combine
%and extending 
multiple 
%sources 
artist modalities into a single representation.
%of artist information. %that are 
%  
%that can be used for the full catalog of artists. 
%Our evaluation shows 
In an artist similarity task, we show our method yields clear improvements 
over other methods
in terms of retrieval accuracy and 
%giving a better distribution of the retrieved artists, 
coverage,
%be more robust to missing modality data.
and successfully combines complementary information from diverse modalities.
In particular, we investigate  retrieval bias towards the query's modality.
%
%We also investigate, for the first time, the bias that a contrastive model may have to different groups of artists to be retrieved depending on the type of data that is available and used for them. 
%
Although our method exhibits a slight bias towards retrieving artists with similar modality to the query, we show it handles cross-modal retrieval better than other methods.
Future work may be dedicated to further mitigate this bias.
%
%Our experiments with an open and an in-house dataset show that in extreme cases of cross-modal retrieval where artists have limited information available, our contrastive method slightly favors artists where only the same information of the query artist is available but it is also able to retrieve artists where only other information is available,  allowing us to recommend more diverse artists. 
Additionally, we show that our method
%our analysis indicates that the contrastive method shows a substantial gain for all the popularity levels of artists, and it 
is particularly beneficial for less popular artists. 

%compared to the baseline models.

%We can see that artists with the same information as the query artist may be slightly favored. We see that in particular when training the model with OWN, if only CF information is available there is a higher tendency to retrieve other artists with only CF information available. Therefore, while the proposed contrastive method allows us to operate in a single common space for all the artists and shows good performance according to precision and coverage, we see a point of improvement in the future to reduce the bias that may generate to the availability of the information in this specific situation.

%We further investigate the embeddings produced by the contrastive method in terms of the coherence and uniqueness of the different modalities for each artist, and identify some aspects that may lead to improvements in future work. 

Our method appears to generate an artist representation space with high local coherence for intra-artist modalities, but at the cost of inter-artist separation.
Depending on the final application, this is a property that could perhaps be managed by iterating on the contrastive learning method, for instance, by adapting the loss function or by adapting the size of the training sample batch as suggested in~\cite{NEURIPS2022_db174d37}.
%
%It would be worthwhile to further study the effect of batch size in the angle of the coherency-uniqueness, which might give us %the further 
%insights of ways to improve the model, since the size of the mini-batch implicitly controls the confidence of the ``negativity'' of %negative 
%the samples and there is evidence that such difference in sampling %``bias'' 
%can affect the training~\cite{NEURIPS2022_db174d37}. 

\section{ACKNOWLEDGEMENT}
We would like to express special thanks to Matt McCallum for the help collecting audio features and Sam Sandberg for his valuable comments.

% For bibtex users:
\bibliography{ISMIRtemplate}

\begin{thebibliography}{10}
\providecommand{\url}[1]{#1}
\csname url@samestyle\endcsname
\providecommand{\newblock}{\relax}
\providecommand{\bibinfo}[2]{#2}
\providecommand{\BIBentrySTDinterwordspacing}{\spaceskip=0pt\relax}
\providecommand{\BIBentryALTinterwordstretchfactor}{4}
\providecommand{\BIBentryALTinterwordspacing}{\spaceskip=\fontdimen2\font plus
\BIBentryALTinterwordstretchfactor\fontdimen3\font minus
  \fontdimen4\font\relax}
\providecommand{\BIBforeignlanguage}[2]{{%
\expandafter\ifx\csname l@#1\endcsname\relax
\typeout{** WARNING: IEEEtran.bst: No hyphenation pattern has been}%
\typeout{** loaded for the language `#1'. Using the pattern for}%
\typeout{** the default language instead.}%
\else
\language=\csname l@#1\endcsname
\fi
#2}}
\providecommand{\BIBdecl}{\relax}
\BIBdecl

\bibitem{ellis2002quest}
D.~P. Ellis, B.~Whitman, A.~Berenzweig, and S.~Lawrence, ``The quest for ground
  truth in musical artist similarity,'' in \emph{ISMIR}, 2002.

\bibitem{pohle2009rhythm}
T.~Pohle, D.~Schnitzer, M.~Schedl, P.~Knees, and G.~Widmer, ``On rhythm and
  general music similarity.'' in \emph{ISMIR}, 2009, pp. 525--530.

\bibitem{schedl2014harvesting}
M.~Schedl, D.~Hauger, and J.~Urbano, ``Harvesting microblogs for contextual
  music similarity estimation: a co-occurrence-based framework,''
  \emph{Multimedia Systems}, vol.~20, pp. 693--705, 2014.

\bibitem{oramas2015semantic}
S.~Oramas, M.~Sordo, L.~Espinosa-Anke, and X.~Serra, ``A semantic-based
  approach for artist similarity,'' in \emph{ISMIR}, 2015.

\bibitem{korzeniowski2022artist}
F.~Korzeniowski, S.~Oramas, and F.~Gouyon, ``Artist similarity for everyone: A
  graph neural network approach,'' \emph{Transactions of the International
  Society for Music Information Retrieval}, vol.~5, no.~1, 2022.

\bibitem{mccallum2022supervised}
M.~C. McCallum, F.~Korzeniowski, S.~Oramas, F.~Gouyon, and A.~F. Ehmann,
  ``Supervised and unsupervised learning of audio representations for music
  understanding,'' in \emph{ISMIR}, 2022.

\bibitem{alonso2020tensorflow}
P.~Alonso-Jim{\'e}nez, D.~Bogdanov, J.~Pons, and X.~Serra, ``Tensorflow audio
  models in {E}ssentia,'' in \emph{IEEE International Conference on Acoustics,
  Speech and Signal Processing (ICASSP)}, 2020, pp. 266--270.

\bibitem{alonso2022music}
P.~Alonso-Jim{\'e}nez, X.~Serra, and D.~Bogdanov, ``Music representation
  learning based on editorial metadata from {D}iscogs,'' in \emph{ISMIR}, 2022.

\bibitem{dieleman2011audio}
S.~Dieleman, P.~Brakel, and B.~Schrauwen, ``Audio-based music classification
  with a pretrained convolutional network,'' in \emph{ISMIR}, 2011, pp.
  669--674.

\bibitem{oramas2018multimodal}
S.~Oramas, F.~Barbieri, O.~Nieto~Caballero, and X.~Serra, ``Multimodal deep
  learning for music genre classification,'' \emph{Transactions of the
  International Society for Music Information Retrieval. 2018; 1 (1): 4-21.},
  2018.

\bibitem{won2021multimodal}
M.~Won, S.~Oramas, O.~Nieto, F.~Gouyon, and X.~Serra, ``Multimodal metric
  learning for tag-based music retrieval,'' in \emph{IEEE International
  Conference on Acoustics, Speech and Signal Processing (ICASSP)}, 2021, pp.
  591--595.

\bibitem{ferraro2019bias}
\BIBentryALTinterwordspacing
A.~Ferraro, ``Music cold-start and long-tail recommendation: Bias in deep
  representations,'' in \emph{Proceedings of the 13th ACM Conference on
  Recommender Systems}, 2019, p. 586–590. [Online]. Available:
  \url{https://doi.org/10.1145/3298689.3347052}
\BIBentrySTDinterwordspacing

\bibitem{oramas2017multi}
S.~Oramas, O.~Nieto, F.~Barbieri, and X.~Serra, ``Multi-label music genre
  classification from audio, text, and images using deep features,'' in
  \emph{ISMIR}, 2017.

\bibitem{van2013deep}
A.~Van~den Oord, S.~Dieleman, and B.~Schrauwen, ``Deep content-based music
  recommendation,'' \emph{Advances in neural information processing systems},
  vol.~26, 2013.

\bibitem{huang2022mulan}
Q.~Huang, A.~Jansen, J.~Lee, R.~Ganti, J.~Y. Li, and D.~P. Ellis, ``Mulan: A
  joint embedding of music audio and natural language,'' in \emph{ISMIR}, 2022.

\bibitem{manco2022contrastive}
I.~Manco, E.~Benetos, E.~Quinton, and G.~Fazekas, ``Contrastive audio-language
  learning for music,'' in \emph{ISMIR}, 2022.

\bibitem{oord2018representation}
A.~v.~d. Oord, Y.~Li, and O.~Vinyals, ``Representation learning with
  contrastive predictive coding,'' \emph{arXiv preprint arXiv:1807.03748},
  2018.

\bibitem{ferraro2021enriched}
A.~Ferraro, X.~Favory, K.~Drossos, Y.~Kim, and D.~Bogdanov, ``Enriched music
  representations with multiple cross-modal contrastive learning,'' \emph{IEEE
  Signal Processing Letters}, vol.~28, pp. 733--737, 2021.

\bibitem{Bertin-Mahieux2011}
T.~Bertin-Mahieux, D.~P. Ellis, B.~Whitman, and P.~Lamere, ``The million song
  dataset,'' in \emph{ISMIR}, 2011.

\bibitem{DBLP:conf/icdm/HuKV08}
Y.~Hu, Y.~Koren, and C.~Volinsky, ``Collaborative filtering for implicit
  feedback datasets,'' in \emph{Proceedings of the 8th {IEEE} International
  Conference on Data Mining {(ICDM} 2008)}, 2008, pp. 263--272.

\bibitem{tim_head_2018_1207017}
\BIBentryALTinterwordspacing
T.~Head, MechCoder, G.~Louppe, I.~Shcherbatyi, fcharras, Z.~Vinícius,
  cmmalone, C.~Schröder, nel215, N.~Campos, T.~Young, S.~Cereda, T.~Fan, rene
  rex, K.~K. Shi, J.~Schwabedal, carlosdanielcsantos, Hvass-Labs, M.~Pak,
  SoManyUsernamesTaken, F.~Callaway, L.~Estève, L.~Besson, M.~Cherti,
  K.~Pfannschmidt, F.~Linzberger, C.~Cauet, A.~Gut, A.~Mueller, and A.~Fabisch,
  ``scikit-optimize/scikit-optimize: v0.5.2,'' Mar. 2018. [Online]. Available:
  \url{https://doi.org/10.5281/zenodo.1207017}
\BIBentrySTDinterwordspacing

\bibitem{korzeniowski2021artist}
F.~Korzeniowski, S.~Oramas, and F.~Gouyon, ``Artist similarity with graph
  neural networks,'' in \emph{ISMIR}, 2021.

\bibitem{DBLP:conf/kdd/BinghamM01}
E.~Bingham and H.~Mannila, ``Random projection in dimensionality reduction:
  applications to image and text data,'' in \emph{Proceedings of the seventh
  {ACM} {SIGKDD} international conference on Knowledge discovery and data
  mining}, 2001, pp. 245--250.

\bibitem{Johnson1984}
\BIBentryALTinterwordspacing
W.~B. Johnson and J.~Lindenstrauss, ``Extensions of {L}ipschitz mappings into a
  {H}ilbert space,'' \emph{Contemporary Mathematics}, vol.~26, pp. 189--206,
  1984. [Online]. Available: \url{https://doi.org/10.1090/conm/026/737400}
\BIBentrySTDinterwordspacing

\bibitem{scikit-learn}
F.~Pedregosa, G.~Varoquaux, A.~Gramfort, V.~Michel, B.~Thirion, O.~Grisel,
  M.~Blondel, P.~Prettenhofer, R.~Weiss, V.~Dubourg, J.~Vanderplas, A.~Passos,
  D.~Cournapeau, M.~Brucher, M.~Perrot, and E.~Duchesnay, ``Scikit-learn:
  Machine learning in {P}ython,'' \emph{Journal of Machine Learning Research},
  vol.~12, pp. 2825--2830, 2011.

\bibitem{valcarce2020assessing}
D.~Valcarce, A.~Bellog{\'\i}n, J.~Parapar, and P.~Castells, ``Assessing ranking
  metrics in top-n recommendation,'' \emph{Information Retrieval Journal},
  vol.~23, pp. 411--448, 2020.

\bibitem{DBLP:books/sp/EfronT93}
\BIBentryALTinterwordspacing
B.~Efron and R.~Tibshirani, \emph{An Introduction to the Bootstrap}.\hskip 1em
  plus 0.5em minus 0.4em\relax Springer, 1993. [Online]. Available:
  \url{https://doi.org/10.1007/978-1-4899-4541-9}
\BIBentrySTDinterwordspacing

\bibitem{NEURIPS2022_db174d37}
C.~Chen, J.~Zhang, Y.~Xu, L.~Chen, J.~Duan, Y.~Chen, S.~Tran, B.~Zeng, and
  T.~Chilimbi, ``Why do we need large batchsizes in contrastive learning? a
  gradient-bias perspective,'' in \emph{Advances in Neural Information
  Processing Systems}, vol.~35, 2022, pp. 33\,860--33\,875.

\end{thebibliography}

% For non bibtex users:
%\begin{thebibliography}{citations}
% \bibitem{Author:17}
% E.~Author and B.~Authour, ``The title of the conference paper,'' in {\em Proc.
% of the Int. Society for Music Information Retrieval Conf.}, (Suzhou, China),
% pp.~111--117, 2017.
%
% \bibitem{Someone:10}
% A.~Someone, B.~Someone, and C.~Someone, ``The title of the journal paper,''
%  {\em Journal of New Music Research}, vol.~A, pp.~111--222, September 2010.
%
% \bibitem{Person:20}
% O.~Person, {\em Title of the Book}.
% \newblock Montr\'{e}al, Canada: McGill-Queen's University Press, 2021.
%
% \bibitem{Person:09}
% F.~Person and S.~Person, ``Title of a chapter this book,'' in {\em A Book
% Containing Delightful Chapters} (A.~G. Editor, ed.), pp.~58--102, Tokyo,
% Japan: The Publisher, 2009.
%
%
%\end{thebibliography}

\end{document}